\pdfoutput=1

\documentclass[11pt]{article}

\usepackage{times,latexsym}
\usepackage{url}
\usepackage[T1]{fontenc}
\usepackage{microtype}
\usepackage{inconsolata}
\usepackage{booktabs}
\usepackage{xspace}
\usepackage{amsmath}
\usepackage{amssymb}
\usepackage{mathtools}
\usepackage{tcolorbox}
\usepackage{caption}
\usepackage{subcaption}
\usepackage[hang,flushmargin]{footmisc}
\usepackage{multirow}
\usepackage{enumitem}
\usepackage{balance}

\newcommand{\rv}{Rank\-Vicuna\xspace}
\newcommand{\rz}{Rank\-Zephyr\xspace}

\newcommand{\rzp}{Rank\-Zephyr\textsubscript{$\rho$}\xspace}
\newcommand{\msvone}{MS MARCO v1 passage ranking task\xspace}
\newcommand{\msvtwo}{MS MARCO v2 passage ranking task\xspace}

\newcommand{\rg}{Rank\-GPT\xspace}

\newcommand{\rgthreefive}{Rank\-GPT\textsubscript{3.5}\xspace}
\newcommand{\rgfour}{Rank\-GPT\textsubscript{4}\xspace}
\newcommand{\repllama}{Rep\-LLaMA\xspace}
\newcommand{\rankllama}{Rank\-LLaMA\xspace}
\newcommand{\gptthreefive}{GPT\textsubscript{3.5}\xspace}
\newcommand{\adatwo}{ADA\textsubscript{2}\xspace}
\newcommand{\gptfour}{GPT\textsubscript{4}\xspace}

\usepackage[]{acl}
\usepackage{xspace,mfirstuc,tabulary}

\title{\rz: Effective and Robust Zero-Shot \\ Listwise Reranking is a Breeze!}

\author{Ronak Pradeep, Sahel Sharifymoghaddam, Jimmy Lin \\[1ex]
David R.\ Cheriton School of Computer Science,\\
University of Waterloo, Canada \\[1ex]
\texttt{\{rpradeep, sahel.sharifymoghaddam, jimmylin\}@uwaterloo.ca}}
\begin{document}
\maketitle

\begin{abstract}
In information retrieval, proprietary large language models (LLMs) such as \gptfour and open-source counterparts such as LLaMA and Vicuna have played a vital role in reranking.
However, the gap between open-source and closed models persists, with reliance on proprietary, non-transparent models constraining reproducibility.
Addressing this gap, we introduce \rz, a state-of-the-art, open-source LLM for listwise zero-shot reranking. 
\rz not only bridges the effectiveness gap with \gptfour but in some cases surpasses the proprietary model.
Our comprehensive evaluations across several datasets (TREC Deep Learning Tracks; NEWS and COVID from BEIR) showcase this ability.
\rz benefits from strategic training choices and is resilient against variations in initial document ordering and the number of documents reranked. 
Additionally, our model outperforms \gptfour on the NovelEval test set, comprising queries and passages past its training period, which addresses concerns about data contamination.
To foster further research in this rapidly evolving field, we provide all code necessary to reproduce our results at \url{https://github.com/castorini/rank_llm}. 
\end{abstract}

\section{Introduction}

The emergence of instruction fine-tuned large language models (LLMs) has expanded the horizon of applications in areas ranging from natural language processing to information retrieval to software engineering.
This development has been particularly influential in the realm of text ranking. 
Notably, there has been a surge in initiatives focused on zero-shot listwise reranking using LLMs, referred to as ``prompt-decoders'', as evidenced by works such as RankGPT~\cite{RankGPT} and LRL~\cite{LRL}.
However, a common limitation in these endeavors is their reliance on proprietary models.
While these models facilitate rapid prototyping and are conveniently accessible via API endpoints, they pose challenges in terms of scientific reproducibility.
This issue is critical both from the standpoint of adhering to the principles of robust scientific methodology and practically in the context of achieving consistent, reliable measurements in experimental evaluations. 

Recently, RankVicuna~\cite{RankVicuna} helped address this pressing need within the academic community for an open-source LLM that can proficiently execute reranking tasks, improving over the much larger proprietary model \rgthreefive.
However, \rv still lags behind the state-of-the-art \rgfour in effectiveness.
Bridging this gap and striving beyond with an open-source model would be of great value to the NLP and IR communities working towards RAG architectures that require high-precision results.

This paper introduces \rz, an open-source LLM for reranking that bridges the gap and, in a few cases, goes beyond \rgfour, the state-of-the-art reranker.
Our investigation is steered by several research questions, informed by comprehensive experiments and analyses:

\begin{itemize}[leftmargin=*]

\item Can an open-source LLM, specifically \rz with only 7B parameters, improve on the listwise reranking effectiveness of much larger proprietary models such as \rgfour in a zero-shot setting?

\item What is the impact on effectiveness of employing multiple reranking passes over the candidate list?

\item Which training choices, including the source of hard negatives, the teacher model, the types of queries exposed, and window sizes used during distillation, are important for instruction fine-tuning a robust and effective listwise reranker?

\item How do the selection of first-stage retrieval model and the number of top retrieved candidates affect the downstream reranking effectiveness of \rz for both in-domain and out-of-domain effectiveness?

\item In what ways do initial orderings of retrieved documents influence the effectiveness of these listwise rerankers?

\item How do reranking models perform on uncontaminated test queries and passages originating after the model training cut-off?

\end{itemize}

\noindent Our comprehensive evaluations encompass both in-domain datasets---TREC 2019 and 2020 Deep Learning Tracks~\cite{dl19,dl20} from the \msvone---as well as out-of-domain datasets---TREC 2021 and 2022 Deep Learning Tracks~\cite{dl21,dl22} from the \msvtwo, NEWS and COVID from the BEIR collection~\cite{beir}, and NovelEval from Google Web Search~\cite{RankGPT}.
These experiments reveal \rz's capability to rival and occasionally surpass the reranking effectiveness of \rgfour.

Remarkably, \rz achieves this feat by building on an open-source LLM with orders of magnitude fewer parameters.
The central insights from our work include:

\begin{itemize}[leftmargin=*]

\item \textbf{Effectiveness of \rz:} While proprietary LLMs such as \rgfour exhibit high reranking effectiveness, they come with limitations like non-reproducibility, non-deterministic outputs, and obscurity behind an API, impeding their utility in academic settings. 
In contrast, \rz is a much smaller open-source model that is competitive in terms of quality with deterministic behavior and is publicly available for researchers to build on.
Additionally, given that the paradigm of listwise prompt decoding with models such as \rz is relatively new, we find that the absence of their results in judgment pools can penalize the measurement of retrieval effectiveness, despite already achieving the state of the art.

\item \textbf{Effect of Progressive Reranking:} Our exploration shows that \rz, when subjected to multiple reranking passes --- a process we dub ``\rzp'' --- generally results in better final effectiveness.

\item \textbf{Effect of Training Data Choices:} Our findings demonstrate that the effectiveness of \rz benefits from strategic choices in training data sources.
Employing \rgfour as the teacher model with a small set of 5K reorderings of \adatwo as the first-stage retrieval model contributes to closing the effectiveness gap with \rgfour. 
Furthermore, instruction distillation with variable window sizes and shuffled input ordering enhances the listwise reranking capability of \rz, making it robust to different reranking scenarios.

\item \textbf{Effect of First-Stage Retrieval Models:} Our experiments show that \rz consistently improves over different first-stage retrieval models and that improved candidate lists generally result in better reranking quality.
We also demonstrate that a more nuanced exploration of different first-stage retrieval methods is required to capture many details, as opposed to a single-source study as seen in prior work~\cite{RankGPT, rankllama}.

\item \textbf{Effect of Initial Document Ordering:} At inference time, the initial ordering of candidate documents influences reranking effectiveness.
However, \rz maintains robustness even when presented with shuffled orderings.

\item \textbf{Effectiveness on Uncontaminated Test Sets:} On NovelEval, a set of uncontaminated test queries and passages, \rz outperforms its proprietary counterparts based on the primary evaluation metric.
This illustrates its generalizability and effectiveness in real-world scenarios where the model tackles evolving information needs on an ever-changing web.

\end{itemize}

\noindent By sharing model checkpoints and associated code, we aim to contribute a valuable resource to the research community. 
Our work provides a foundation for future explorations in developing more effective and efficient reranking models, addressing the growing demand for high-quality retrieval systems in the era of retrieval-augmented LLMs.

\section{Background and Related Work}

The core task governing information retrieval is: given a document corpus $\mathcal{C}=\{D_1, D_2, \ldots, D_n\}$ and a query $q$, return a ranked list of $k$ highly relevant documents from $\mathcal{C}$, where $k \ll |\mathcal{C}|$. 
The notion of relevance is determined by metrics such as nDCG or average precision. 
Subsequently, the role of a \textit{reranker} is to enhance the quality of this ranked list, either using the same or different metrics, typically by refining the selection made by the retriever or a preceding reranker.

Both retrievers and rerankers are integral components of multi-stage pipelines for text ranking. 
While recent studies have focused on transformers~\cite{duobert, lce}, the general design traces back over a decade~\cite{Matveeva_etal_SIGIR2006, Cambazoglu_etal_WSDM2010, Wang_etal_SIGIR2011}. 
In the transformer era, a notable milestone was the introduction of monoBERT by \citet{monobert}, employing an encoder-only transformer model (BERT) for reranking in a cross-encoder setup.
This marked a significant leap in reranking effectiveness over previous approaches~\cite{Lin_etal_2021_ptr4tr}.
Afterwards, research expanded to include reranking using decoder-only models~\cite{nogueira-dos-santos-etal-2020-beyond, rankllama} and encoder--decoder models~\cite{monoT5,rankt5}. 
Although these models are effective, they rely on fine-tuning with extensive training data, often using the \msvone~\citep{msmarco}. 

Initially, transformer-based rerankers focused on a pointwise approach, assessing document relevance in isolation.
However, recent advancements have adopted pairwise and listwise losses in cross-encoder models~\cite{lce, SqueezeMBERT, rankt5}, enhancing reranking accuracy, particularly in high-precision contexts, and achieving better alignment with initial retrievers. 
These methods often incorporate hard negatives to improve system effectiveness.

In the pairwise paradigm, reranking models estimate whether one document is \emph{more relevant} than another document for a particular query.
\citet{duobert} introduced pairwise reranking with encoder-only models, duoBERT, and \citet{EMD} further expanded on this idea with duoT5, an effective pairwise reranker using encoder--decoder models.
Once pairwise scores are computed by the model, they are aggregated to assign a single score for each document.

The listwise paradigm generalizes the pairwise paradigm to directly reorder a list of input documents in terms of relevance.
Our study focuses on zero-shot scenarios, where no task-specific training data, such as relevant query--passage pairs, is provided.
This approach builds on recent advances using large language models as rerankers in multi-stage pipelines, mainly leveraging prompt engineering~\cite{RankGPT, LRL, PRP, zhuang2023setwise, zhuang2023beyond}.

More specifically, we build on the success of the recent \rv model~\cite{RankVicuna} that leverages instruction fine-tuning to distill the listwise reranking effectiveness of \rgthreefive~\cite{RankGPT} down to the 7B parameter Vicuna model, a LLaMA variant.
The authors introduced the term ``prompt-decoders'', contrasting with BERT-style cross-encoders, to describe LLM-based rerankers operating in a zero-shot fashion. 
While \rv improved on its teacher \rgthreefive, a gap between the model and the state-of-the-art, \rgfour still exists.
Additionally, \rv was not trained to handle a variable number of input documents, leading to the inability of the model to generalize to an arbitrary number of candidates within the maximum input token size.
Our work addresses these shortcomings.

In our experiments, the initial candidate document lists are sourced from various first-stage retrievers, traditional and neural, supervised and unsupervised, sparse and dense, to examine their downstream effects.
While various retrieval techniques exist (sparse, dense, hybrid), our focus is not on them but rather on their effect on reranking.
We also do not explore the effect of reranking candidates from other rerankers, as seen in~\citet{EMD, vera, CT22, naverlootrec23}

Another related area of research involves using LLMs to create synthetic queries for generating relevant training pairs for retrievers or rerankers, as seen in InPars \citep{inpars, inpars-light}, Promptagator \citep{promptagator}, and HyDE \citep{hyde}. 
\citet{DSIScaling} explored generative retrieval, emphasizing the necessity of synthetic queries for effectiveness with larger corpora. 
These methods, however, utilize LLMs indirectly.
Here, we not only generate instruction data from RankGPT but also \emph{directly} instruction fine-tune the LLM, which is Zephyr\textsubscript{$\beta$} in our case.

\section{\rz}

\subsection{Prompt Design}
\label{sec:prompt}

This study builds on recent advances in LLM-based reranking methodologies, particularly those employing zero-shot \textit{listwise} approaches~\citep{LRL, RankGPT}. 
The superiority of \textit{listwise} reranking lies in its capacity to concurrently perform relevance comparisons over multiple documents to determine the best final reordering.
Our work structures the reranking problem in the following manner: 
Given a user query denoted as $q$ and a set of candidate documents $\{D_1, \ldots, D_n\}$ from a previous retrieval stage, the objective is to reorder these documents into a list that optimizes a specific retrieval metric such as normalized Discounted Cumulative Gain (nDCG).

To implement zero-shot listwise reranking, we employ a prompt template analogous to the \rg prompt outlined by~\citet{RankGPT} and further refined by~\citet{RankVicuna}.
The system description and structure are adapted to accommodate variances between the prompt templates.
\autoref{prompt:1} shows the precise input prompt utilized for \rz.
Unlike in \rv, we prepend the actual reranking prompt with a generic system description that would work for any listwise reranking LLM.
We hope that aligning our model with the exact prompt setup used to train the underlying LLM would result in faster and more effective instruction distillation from the teacher.

\begin{figure}[t]
\begin{small}
\begin{verbatim}
<|system|>
You are RankLLM, an intelligent 
assistant that can rank passages based on 
their relevancy to the query.
<|user|>
I will provide you with {num} passages,
each indicated by a numerical identifier [].
Rank the passages based on their relevance to
the search query: {query}.

[1] {passage 1}
[2] {passage 2}
...
[{num}] {passage {num}}

Search Query: {query}.

Rank the {num} passages above based on their 
relevance to the search query. All the
passages should be included and listed
using identifiers, in descending order of
relevance. The output format should be [] > [],
e.g., [4] > [2]. Only respond with the ranking
results, do not say any word or explain.
<|assistant|>
\end{verbatim}
\emph{Model 
Generation: [9] > [4] > [20] > \ldots > [13]}
\end{small}
\caption{The reranking prompt and sample generation for \rz.}
\label{prompt:1}
\end{figure}

\subsection{Training Stage 1: Distilling from \gptthreefive}
\label{sec:rankzephyr-stage1}

Our approach began with an initial training phase using ranked lists produced by \rgthreefive. 
These lists were based on a set of 100K training queries sourced from the MS MARCO v1 passage ranking training dataset and provided by \citet{RankGPT}. 
For each query, the top 20 candidate passages were retrieved using BM25 with Pyserini~\cite{Pyserini}, which were then ranked by \rgthreefive to create the teacher reorderings that were subsequently distilled into our student model, \rz.
Note that neither model was trained with human-annotated query--passage relevance pairs, thus keeping our approach zero-shot.

To enhance the quality and robustness of our models, we incorporated additional steps as outlined by \citet{RankVicuna}:

\begin{itemize}[leftmargin=*]

\item We excluded malformed generations from our training data. These include examples of improper list formatting, missing document identifiers, or repetitive elements.

\item We included not only the original rankings from the teacher model, which reorders the top 20 BM25 results~\citep{bm25}, but also introduced shuffled input orders. 
Our intention is to present a more challenging reordering task to the model without additional data generation. 
Nonetheless, the original BM25 orderings were also included to emulate a scenario closest to model inference.

\end{itemize}

\noindent We trained the 7B parameter Zephyr\textsubscript{$\beta$} model~\cite{zephyr}, built on Mistral~\cite{Mistral}, for three epochs with an effective batch size of 64 and a learning rate of $5 \times 10^{-6}$, using bfloat16 precision using \rgthreefive as the teacher (as described above).
Model training used the \texttt{axolotl} library,\footnote{\url{https://github.com/OpenAccess-AI-Collective/axolotl}} and leveraged noisy embeddings, shown to improve instruction fine-tuning~\cite{neft}.
The training process took approximately 40 hours on eight NVIDIA RTX A6000 GPUs. 
Given the observed effectiveness gap between \rgthreefive and \rgfour, this work focuses on further fine-tuning our \rz model using \rgfour data, described below.

\subsection{Training Stage 2: Distilling from \gptfour}
\label{sec:rankzephyr-stage2}

Due to the prohibitive cost of generating high-quality \rgfour training data for the entire set of 100K queries, estimated at over \$10K USD, we limited our experimentation to a subset of no more than 5K queries. 
The selection methods for these queries are as follows:
\begin{itemize}[leftmargin=*]
    \item \textbf{Random:} Queries are chosen randomly from the 100K query set curated by~\citet{RankGPT}.
    \item \textbf{Discriminative:} Queries are iteratively selected based on their distinctiveness. 
We do this by choosing a query from the remaining set whose \adatwo embedding minimizes the maximum inner product score with the embeddings of the already selected queries, promoting diversity in the training samples.
\end{itemize}

\noindent In our quest to surpass the effectiveness of \rgfour, we also investigated the effects of substituting BM25 candidate orderings with those generated by \adatwo; this is a zero-shot embedding model and thus the approach maintains the zero-shot nature of entire pipeline.
We hope this exposes our model to teacher reorderings over a ``harder'' set, akin to how hard negatives help improve the effectiveness of cross-encoder models~\cite{lce, SqueezeMBERT, rankt5}.

Additionally, we addressed a limitation in \rv related to its fixed training window size and the subsequent inference limitations over smaller window sizes.
By training our model on a subset of $p$ ($\leq 20$) passages, randomly chosen from the original input and ordered by the teacher model, we exposed \rz to variable-length reranking tasks. 
This strategy also has the benefit of not requiring further \rgfour querying for teacher reordering.
Unless mentioned otherwise, we employed this strategy and sampled 3 subsets of the 20 passages to augment our training set.

To be clear, the second-phase fine-tuning leveraging \rgfour as the teacher started with an already effective model following the basic \rv recipe, but on Zephyr\textsubscript{$\beta$} instead (as described in Section~\ref{sec:rankzephyr-stage1}).
This two-phase approach drastically reduces the additional training time.
The second-phase training consumed about a tenth of the effort of fine-tuning from the much larger \rgthreefive dataset.

\subsection{Inference}

Reranking with \rz is straightforward, similar to \rv.
Candidates from the first-stage retriever are fed into the prompt-decoder to obtain the final ranked list.

To rerank the top 100 candidates, we employed a sliding window approach akin to \rg and \rv. 
In our trials, we maintained the same parameters as \rg (window size of 20, stride of 10) to isolate its impact in our comparative analysis unless explicitly stated otherwise.

To align with the \rv evaluation protocol, all our reranking experiments substituted any instance of the format $[n]$ in passages with $(n)$ to prevent model confusion. 
Additionally, we utilized the \texttt{fix\_text} function from \texttt{ftfy} to preprocess any input to the rerankers.

\citet{RankVicuna} found that progressive reranking with multiple sliding passes (up to three times) helped further refine the ranked list.
That is, the resulting ranked list of the first pass serves as the input to the second pass, and so on---so the passes happen \emph{sequentially}.
In light of this previous result, we introduced a setting, \rzp, wherein the \rz model progressively reranks the input three times.

\begin{table*}[t]
\centering \scalebox{0.75}{
\begin{tabular}{l|llllllll}
\toprule
\toprule
\multicolumn{1}{l}{\multirow{2}{*}{\hspace{100pt}}} & 
\multicolumn{2}{l}
{\multirow{2}{*}
{\begin{tabular}
[c]{@{}c@{}}\textbf{Source} \\
\textbf{Prev.} \hspace{35pt}  \textbf{Top-$k$}
\end{tabular}}} &
\multicolumn{2}{l}
{\multirow{2}{*}
{\begin{tabular}
[c]{@{}c@{}}\textbf{MSv1} \\
\textbf{DL19} \hspace{8pt}  \textbf{DL20}
\end{tabular}}} &
\multicolumn{2}{l}
{\multirow{2}{*}
{\begin{tabular}
[c]{@{}c@{}}\textbf{MSv2} \\
\textbf{DL21} \hspace{8pt}  \textbf{DL22}
\end{tabular}}} &
\multicolumn{2}{l}
{\multirow{2}{*}
{\begin{tabular}
[c]{@{}c@{}}\textbf{BEIR} \\
\textbf{NEWS} \hspace{4pt}  \textbf{COVID}
\end{tabular}}}\\
\multicolumn{1}{l}{}\ & \multicolumn{1}{l}{}\ & \multicolumn{2}{c}{} & \multicolumn{2}{c}{} \\
\midrule
(1) BM25                                        &
\multicolumn{2}{l}{None}{$|C|$}                 &
0.5058 & 0.4796 &  0.4458 & 0.2692 & 0.3952 & 0.5947                             \\
(2) SPLADE++ ED &
\multicolumn{2}{l}{None}{$|C|$}            &
0.7308                                &
0.7195  & 0.6846 & 0.5705 & 0.4152 & 0.7274\\
\midrule
(3) \rankllama-13B &
\multicolumn{2}{l}{\repllama}{100} & 0.7567 & 0.7748 & - & - & - & - \\
\midrule
(4) \rgfour                  &
\multicolumn{2}{l}{SPLADE++ ED}{100}                   &
0.7464       &
0.7076  & 0.7721 & 0.7175 & 0.5327 & 0.8792   \\
(5) \rgfour (PSC)                    &
\multicolumn{2}{l}{SPLADE++ ED}{100} &
0.7601        &
0.7514      & - & - & - & -    \\
\midrule
(6) \rv & 
\multicolumn{2}{l}{SPLADE++ ED}{100}            & 
0.7459 & 0.7473 &   0.7011 & 0.5817 & 0.4750 & 0.8283                          \\
\midrule

(7) \rz & \multicolumn{2}{l}{SPLADE++ ED}{100} &
0.7816  &
0.8159  & 0.7598 & 0.6692 & 0.5060 & 0.8535  \\
(8) \rzp & \multicolumn{2}{l}{SPLADE++ ED}{100} &
0.7855  &
0.8255 & 0.7688  &  0.6628 &  0.5107 & 0.8566 \\
\bottomrule 
\bottomrule                                            
\end{tabular}}
\caption{nDCG@10  on DL19 and DL20 from the \msvone (MSv1), DL21 and DL22 from the \msvtwo (MSv2), and NEWS and COVID from the BEIR collection for different reranking pipelines.
Each reranker uses the top 100 retrieved results of the previous stage as input. 
Row~(5) is as reported by \citet{tang2023found}.}
\label{tab:1}
\end{table*}

\section{Experimental Framework}

\textbf{Baselines:} We evaluated \rz with established unsupervised sparse/dense ranking models (BM25 and \adatwo), supervised sparse/dense ranking models  (SPLADE++ ED and \repllama) and proprietary prompt-decoder models:\ \rg~\cite{RankGPT} --- both \rgthreefive and \rgfour variants.
Specifically, \gptthreefive refers to the \texttt{gpt-3.5-turbo} model from OpenAI, whereas \gptfour corresponds to \texttt{gpt-4}.

\medskip
\noindent \textbf{Datasets:} Our evaluation leveraged test collections from the TREC 2019 and 2020 Deep Learning Tracks~\cite{dl19, dl20}, incorporating queries and relevance assessments from the passage retrieval task. 
These collections employed the MS MARCO v1 passage corpus~\citep{msmarco}, consisting of approximately 8.8 million passages.
Further, to assess \rz's adaptability to various tasks beyond those using the MS MARCO v1 dataset, we conducted evaluations on two additional sets of collections:
(1) TREC 2021 and 2022 Deep Learning Tracks~\cite{dl21, dl22}, which utilize the MS MARCO v2 passage corpus containing around 138 million passages;
(2) the NEWS and COVID collections from BEIR~\cite{beir}. 
For convenience, we refer to these evaluation sets as DL19--DL22, NEWS, and COVID.

\medskip
\noindent \textbf{Evaluation Choices:}
Unless otherwise specified, \rz corresponds to the reranking model that was first trained on all of the \rgthreefive data, followed by further training on the \rgfour data using 5K queries, sampled randomly, with input orderings from \adatwo.
\rzp corresponds to the same model, but with progressive reranking of the candidate list three times.

The context size was set to $4096$ for \gptthreefive and Zephyr, and $8192$ for \gptfour. 
Due to the non-deterministic outputs of \gptthreefive and \gptfour (even at a zero temperature setting), as noted by~\citet{RankVicuna}, we present results averaged over six and three runs for \rgthreefive and \rgfour, respectively.
We quantify effectiveness using normalized discounted cumulative gain at a rank cutoff of 10 (nDCG@10) and mean average precision at a rank cutoff of 100 (MAP@100).

\begin{table*}[t]
\centering \scalebox{0.75}{
\begin{tabular}{llcccccc}
\toprule
\toprule
\multicolumn{1}{l}{\multirow{2}{*}{\hspace{80pt}}} & 
\multicolumn{1}{c}{\multirow{2}{*}{\textbf{Teacher}}} & 
\multicolumn{2}{c}{\multirow{2}{*}{\begin{tabular}[c]{@{}c@{}}\textbf{Training Source} \\ \textbf{Prev.} \hspace{12pt}  \textbf{Disc.}\end{tabular}}} & 
\multicolumn{2}{c}{\multirow{2}{*}{\begin{tabular}[c]{@{}c@{}}\textbf{DL19} \\ \textbf{BM25} \hspace{6pt}  \textbf{\adatwo}\end{tabular}}} &
\multicolumn{2}{c}{\multirow{2}{*}{\begin{tabular}[c]{@{}c@{}}\textbf{DL20} \\ \textbf{BM25} \hspace{6pt}  \textbf{\adatwo}\end{tabular}}} \\
\multicolumn{1}{l}{}\ & \multicolumn{1}{l}{}\ & \multicolumn{2}{c}{} & \multicolumn{2}{c}{} \\
\midrule
(1) \rv & \rgthreefive & BM25 & $\times$ & 0.6682 & 0.7374 & 0.6549 &  0.7210 \\
\midrule
(2) \rz & \rgthreefive & BM25 & $\times$ & 0.7078 & 0.7161 & 0.6687 & 0.7008 \\
\midrule
(3a) \rz & \rgfour & BM25 & $\times$ & 0.7374 & 0.7645 & 0.6924 & 0.7503 \\
(3b) \rz & \rgfour & BM25 & $\checkmark$ & 0.7275 & 0.7461 & 0.6851 & 0.7550 \\
\midrule
(4a) \rz & \rgfour & \adatwo & $\times$ & 0.7331 & 0.7627 & 0.7185 & 0.7762 \\
(4b) \rz & \rgfour & \adatwo & $\checkmark$  & 0.7248 & 0.7513 & 0.7144 & 0.7779 \\
\midrule
(5) \rz & \rgfour & \adatwo & $\cup$ & 0.7364 & 0.7502 & 0.6941 & 0.7677 \\
\bottomrule 
\bottomrule                                            
\end{tabular}}
\caption{
nDCG@10 on DL19 and DL20 for different reranking methods, with \rgthreefive or \rgfour as the teacher, and BM25 or \adatwo as the training source. The method to select queries (Disc.)\ can be random ($\times$), discriminative ($\checkmark$), or both ($\cup$). Each reranker uses the top 100 retrieved results of the previous stage as input.
Rows (3--5) use the model corresponding to row (2) as the base model. Rows (3--4) are fine-tuned on a set of 1000 queries, while row (5) is fine-tuned on a combination of (4a) and (4b).
Row (1) is as reported by~\citet{RankVicuna}. All these models are trained \emph{only} on reordering 20 passages.
}
\label{tab:ablation_training}
\end{table*}

\section{Results}

\autoref{tab:1} compares different reranking pipelines using DL19 and DL20 data from the \msvone, DL21 and DL22 data from the \msvtwo, and NEWS and COVID from the BEIR collection.
Rows~(1--2) report baselines using only first-stage retrievers:\ BM25~\cite{bm25} and SPLADE++ Ensemble\-Distil (ED)~\cite{spladepp}.
The first represents a traditional ``bag-of-words'' retrieval baseline, while the second is a supervised neural sparse method that has proven effective in a wide range of tasks.

Row~(3) corresponds to RankLLaMA-13B~\cite{rankllama}, a state-of-the-art fine-tuned reranker for the \msvone, reranking the top 100 candidates retrieved by \repllama, a neural dense method, both of which were proposed by~\citet{rankllama}.
It is worth clarifying that despite the name similarities, Rank\-LLaMA is a \emph{pointwise} reranker.

The remaining rows report the effectiveness of using LLM prompt-decoders to reorder the top 100 candidate documents retrieved by SPLADE++ ED.
These include a mix of proprietary models, rows~(4--5), and open-source models, rows~(6--8).
Row~(5) corresponds to the Permutation Self-Consistency method~\cite{tang2023found} employed to aggregate multiple \rgfour results.

Focusing on the \msvone, with the \rz model, we establish state-of-the-art results: 

\begin{itemize}[leftmargin=*]
    \item \rz improves over the highly effective pointwise RankLLaMA-13B, despite not requiring human labels and having only 7B parameters, rows~(7) vs.\ (3). 
    This holds even when we rerank the \repllama candidate set; see row~(3c) in \autoref{tab:first_stages} for more details.

    \item \rz improves over its proprietary \rgfour teacher, rumored to be over two orders of magnitude larger, rows~(7) vs.\ (4).
    The effectiveness boost with \rz is considerably higher in cases where we are processing higher-quality candidate lists from \repllama (more in Section~\ref{sec:ablation_fstage}) or SPLADE++ ED, even seeing relative gains as high as $15\%$. 
    Our method also outshines PSC-aggregated \rgfour, row~(5), which further improves \rgfour by marginalizing out different list orders in the prompt to aggregate a final ranked list.
\end{itemize}

\noindent These results demonstrate the superiority of our open-source, zero-shot \rz model as an effective and robust reranker.
We improve considerably upon \rv, row~(6), the prior state of the art for open-source prompt-decoders.

Additionally, we find that progressive reranking, shown in row~(8), generally improves effectiveness compared to the single-pass counterpart, row~(7), leading to our best results.
This achieves the state of the art for DL19 and DL20, whether open-source or proprietary.
With more refined multi-stage ranking pipelines~\cite{naverlootrec23} feeding even better candidates to \rzp, we would expect even higher scores.
Note that our best system, \rzp, row~(9), has Judged@10 rates of $0.942$ and $0.983$ for DL19 and DL20, respectively.
This means that the nDCG@10 scores reported here represent a lower bound of what such a run would score if part of the judgment pool.

On the \msvtwo and NEWS and COVID tasks from BEIR, whose corpora and queries were never exposed during training, we find that \rz and \rzp remain very effective; here, progressive ranking also helps, rows~(8) vs.\ (7).
Both methods bring huge improvements over their first-stage retriever, SPLADE++ ED, row~(2), and \rv.

However, on these datasets, our models still underperform their \rgfour teacher when reranking SPLADE++ ED output.
Interestingly, though, when reranking lower-quality candidates such as BM25, our models exhibit higher effectiveness than \rgfour (see discussion in Section~\ref{sec:ablation_fstage}).
Nonetheless, this observation helps us understand the current limitations of our models and shows that more work is needed before we have models that are consistently better than \rgfour.

We present two plausible explanations for this gap:\
First, \rz is instruction fine-tuned only on data with a context size of $4096$, as opposed to the $8192$ context of \rgfour, which allows it to better attend to longer biomedical abstracts and news articles in BEIR.
Second, we note 2--4 points lower Judged@10 scores for \rz compared to \rgfour.
Despite all this, our model marks a considerable push in the direction of state-of-the-art generalizable rerankers.

\section{Ablation Studies} 
\begin{table*}[t]
\centering \scalebox{0.75}{
\begin{tabular}{lc|cccccc}
\toprule
\toprule
 & \textbf{Var. WS}  & \textbf{Sliding Window}  & \textbf{OK} & \textbf{Wrong Format} & \textbf{Repetition} & \textbf{Missing} \\
\midrule 

(1) \rgfour & - & 20/10 &  95.11\% & 4.66\% & 0.19& 0.04\%    \\
\midrule
(2a) \rz & $\times$ & 2/1  & 99.91\% & 0.00\% & 0.01\% & 0.08\% \\
(2b) \rz & $\times$ & 10/5  & 38.74\% & 0.98\% & 12.05\% & 48.24\% \\
(2c) \rz & $\times$ & 20/10 & 98.86\% & 0.23\% & 0.80\% & 0.11\% \\ 

\midrule
(3a) \rz & $\checkmark$ & 2/1 & 100.00\% & 0.00\% & 0.00\% & 0.00\% \\
(3b) \rz & $\checkmark$ & 10/5 & 100.00\% & 0.00\% & 0.00\% & 0.00\% \\
(3c) \rz & $\checkmark$ & 20/10  & 99.78\% & 0.11\% & 0.11\% & 0.00\% \\
\bottomrule
\bottomrule
\end{tabular}}
\caption{Distribution of malformed responses over different sliding window choices for \rgfour and \rz. 
Each model reranks the top 100 retrieved results from BM25.  
The Var.\ WS column indicates whether \rz was fine-tuned with variable window sizes. 
Window size of $x/y$ indicates that we evaluate with a sliding window size of $x$ and a stride of $y$.
}
\label{tab:correctness}
\end{table*}

\subsection{Teacher and Training Sources}
\label{sec:ablation_teacher}

Section~\ref{sec:rankzephyr-stage2} discusses several decisions related to fine-tuning with \rgfour data. 
These include query selection methods --- random, discriminative, or a combination of both, and the integration of hard negatives into the prompt decoding paradigm.
\autoref{tab:ablation_training} investigates these choices and their impact on DL19 and DL20 effectiveness using two first-stage retrieval models:\ BM25 and \adatwo.

Row~(1) shows the \rv baseline from \citet{RankVicuna}, which leveraged \rgthreefive reorderings of the BM25 ranked lists of 100K MS MARCO v1 training queries.
Row~(2) examines the effect of moving from Vicuna to Zephyr, which is of comparable parameter count.
Note that for DL19 and DL20, when reranking BM25 candidates, \rz demonstrates higher effectiveness, but when reranking \adatwo candidates, \rv scores higher.
\rz scores higher on average and Zephyr is overall a more capable LLM; these observations provide the rationale for our work to build on Zephyr and not Vicuna.

Rows~(3--5) show the effectiveness of the model corresponding to row (2) fine-tuned on \rgfour data; rows~(3--4) use 1K queries and row~(5) uses 2K queries. 
We see that using \adatwo orderings as the input instead of BM25 generally results in better DL20 effectiveness while having comparable DL19 scores, rows~(4*) vs.\ (3*).
This is expected, as the top 20 \adatwo orderings surpass those of BM25, providing \rz with more sophisticated reorderings from the \rgfour teacher.

Discriminative sampling generally leads to marginally reduced effectiveness, shown in rows (\emph{*b}), compared to rows (\emph{*a}).
This observation indicates that more than covering tail queries, exposing the model to a general distribution of queries and teacher reorderings is important.

To investigate whether combining various query selection strategies is beneficial, we amalgamated the query sets from rows~(4a) and (4b) along with their respective input and teacher reorderings, shown in row (5). 
We observe no notable improvements in effectiveness (despite using more queries). 
Consequently, we decided to expand the method described in row~(4a) --- employing \rgfour reorderings of \adatwo input orderings for a randomly chosen training query set comprising 5K examples.
Unless explicitly stated otherwise, we refer to this setting as the ``full'' \rz model.

\begin{table*}[t]
\centering \scalebox{0.75}{
\begin{tabular}{lccllllll}
\toprule
\toprule
\multicolumn{1}{l}{\multirow{2}{*}{\hspace{80pt}}} & 
\multicolumn{1}{c}{\multirow{2}{*}{\textbf{Var. WS}}} & 
\multicolumn{1}{c}{\multirow{2}{*}{\textbf{Prev.}}} &
\multicolumn{3}{l}
{\multirow{2}{*}
{\begin{tabular}
[c]{@{}c@{}}\textbf{DL19} \\
\hspace{5pt}
\textbf{2/1} \hspace{18pt}    \textbf{10/5} \hspace{18pt}  \textbf{20/10}
\end{tabular}}} &
\multicolumn{3}{l}
{\multirow{2}{*}
{\begin{tabular}
[c]{@{}c@{}}\textbf{DL20} \\
\hspace{5pt}
\textbf{2/1} \hspace{18pt}    \textbf{10/5} \hspace{18pt}  \textbf{20/10}
\end{tabular}}} \\
\multicolumn{1}{l}{}\ & \multicolumn{1}{l}{}\ & \multicolumn{2}{c}{} & \multicolumn{2}{c}{} \\
\midrule
(1a) \rz & $\times$ & BM25 & 0.5771 & 0.6957 & 0.7397 & 0.5700 & 0.6740 & 0.7141 \\
(1b) \rz & $\times$ & \repllama & 0.7593 & 0.7659 & 0.7638 & 0.7495 & 0.7804 & 0.7913 \\
(1c) \rz & $\times$ & SPLADE++ ED & 0.7475 & 0.7703 & 0.7597 & 0.7580 & 0.7970 & 0.8015 \\
\midrule
(2a) \rz & $\checkmark$ & BM25 & 0.5836 & 0.7015 & 0.7420 & 0.5732 & 0.6822 & 0.7086 \\
(2b) \rz & $\checkmark$  &  \repllama & 0.7533 & 0.7694 & 0.7660 & 0.7458 & 0.7715 & 0.7802 \\
(2c) \rz & $\checkmark$  & SPLADE++ ED & 0.7516 & 0.7754 & 0.7816 & 0.7653 & 0.7875 & 0.8159 \\
\bottomrule 
\bottomrule
\end{tabular}}
\caption{nDCG@10 on DL19 and DL20 for different sliding window choices, with BM25, \repllama, and SPLADE++ ED as the first-stage retriever.
Each reranker uses the top 100 retrieved results of the previous stage as input. 
Var.\ WS column indicates whether the \rz was fine-tuned with variable window sizes. 
A window size of $x/y$ indicates that we evaluate with a sliding window size of $x$ and a stride of $y$.}
\label{tab:window_sizes}
\end{table*}
\subsection{Variable Window Listwise Reranking}

In this section, we address a limitation identified in prior work~\cite{RankVicuna}:\ model correctness and effectiveness in the context of a single window size.
Table~\ref{tab:correctness} illustrates the percentage of malformed outputs produced by \rgfour and \rz for different window settings. 
We have grouped the types of system responses following \citet{RankVicuna}:

\begin{enumerate}[leftmargin=*]
\item \textit{Wrong Format}: This category encompasses outputs that fail to adhere to the specified format. For example, \rgfour sometimes refuses to generate a sorted list.
\item \textit{Repetition}: This includes instances where responses exhibit redundant document identifiers.
\item \textit{Missing}: This category accounts for responses with missing document identifiers.
\end{enumerate}

\noindent For \rz, we report results of a single run since our model generates deterministic output:\ identical queries and candidate lists lead to the same exact response. 
Across the sliding window variants, for almost every request in its run, \rz, when trained with sampling, row~(3*), successfully generated correctly formatted responses. 
\rz trained only on windows of size 20 fails to generalize well to sliding windows of size 10, row~(2*).
For \rgfour, we averaged the results of three runs.
Despite being a much larger and more effective general-purpose LLM, \rgfour occasionally returned malformed responses; most of these are refusals to rank.

\autoref{tab:window_sizes} presents the effectiveness of the same two \rz variants with different sliding window configurations on DL19 and DL20, reranking BM25, \repllama, and SPLADE++ ED. 
Rows (1a--c) detail the outcomes of training with a fixed window size, while rows (2a--c) incorporate variable window sizing in training.

In the fixed window size training, RankZephyr demonstrates the general trend of increasing effectiveness, particularly noticeable when reranking candidates from a weaker first-stage retrieval model such as BM25, progressing from 2/1 to 20/10 sliding window configurations on DL19 and DL20.
In a single pass, pairwise approaches may elevate a poorly ranked relevant document to a high position, but they usually fail to reorder multiple items effectively.
On the other hand, listwise methods, with larger context windows, excel at concurrently promoting several poorly ranked relevant documents into higher positions. 

The introduction of variable window sizes generally enhances effectiveness across all first-stage retrievers and window configurations, as evidenced by the higher scores on average, rows~(2a--c). 
Given these observations, the remainder of our experiments leverages the variable window size approach.

\begin{table*}[t]
\centering \scalebox{0.75}{
\begin{tabular}{lllllll}
\toprule
\toprule
\multicolumn{1}{l}{\multirow{2}{*}{\hspace{140pt}}} & 
\multicolumn{2}{l}
{\multirow{2}{*}
{\begin{tabular}
[c]{@{}c@{}}\textbf{Source} \\
\textbf{Prev.} \hspace{70pt}  \textbf{Top-$k$}
\end{tabular}}} &
\multicolumn{2}{l}
{\multirow{2}{*}
{\begin{tabular}
[c]{@{}c@{}}\textbf{DL19} \\
\textbf{nDCG@10} \hspace{30pt}  \textbf{MAP@100}
\end{tabular}}} &
\multicolumn{2}{l}
{\multirow{2}{*}
{\begin{tabular}
[c]{@{}c@{}}\textbf{DL20} \\
\textbf{nDCG@10} \hspace{30pt}  \textbf{MAP@100}
\end{tabular}}} \\
\multicolumn{1}{l}{}\ & \multicolumn{1}{l}{}\ & \multicolumn{2}{c}{} & \multicolumn{2}{c}{} \\
\midrule
(1a) BM25                                       &
\multicolumn{2}{l}{None}{$|C|$}                 &
0.5058 & \hspace{42pt} 0.2476                   &
0.4796 & \hspace{42pt} 0.2685                   \\
(1b) \rz &
\multicolumn{2}{l}{BM25}{20} &
0.6489 & \hspace{42pt} 0.3060 &
0.6196 & \hspace{42pt} 0.3358 \\
(1c) \rz &
\multicolumn{2}{l}{BM25}{100} &
0.7420 & \hspace{42pt} 0.3683 &
0.7086 & \hspace{42pt} 0.4151 \\
\midrule
(2a) SPLADE++ ED                                &
\multicolumn{2}{l}{None}{$|C|$}                 &
0.7308 & \hspace{42pt} 0.4464                   &
0.7197 & \hspace{42pt} 0.4826                   \\
(2b) \rz &
\multicolumn{2}{l}{SPLADE++ ED}{20} &
0.7780 & \hspace{42pt} 0.4691 &
0.7900 & \hspace{42pt} 0.5389 \\
(2c) \rz &
\multicolumn{2}{l}{SPLADE++ ED}{100} &
0.7816 & \hspace{42pt} 0.4908 &
0.8159 & \hspace{42pt} 0.5754 \\
\midrule
(3a) \repllama & \multicolumn{2}{l}{None}{$|C|$}   &
0.7384 & \hspace{42pt} 0.4706                                 &
0.7195 & \hspace{42pt} 0.4927 \\
(3b) \rz &
\multicolumn{2}{l}{\repllama}{20} &
0.7848 & \hspace{42pt} 0.4785 &
0.7799 & \hspace{42pt} 0.5274 \\
(3c) \rz &
\multicolumn{2}{l}{\repllama}{100} &
0.7660 & \hspace{42pt} 0.4887 &
0.7802 & \hspace{42pt} 0.5545 \\
\midrule
(4a) \adatwo                                &
\multicolumn{2}{l}{None}{$|C|$}                 &
0.7035 & \hspace{42pt} 0.4151                   &
0.6759 & \hspace{42pt} 0.4587                   \\
(4b) \rz &
\multicolumn{2}{l}{\adatwo}{20} &
0.7633 & \hspace{42pt} 0.4428 &
0.7568 & \hspace{42pt} 0.5024 \\
(4c) \rz &
\multicolumn{2}{l}{\adatwo}{100} &
0.7573 & \hspace{42pt} 0.4632 &
0.7757 & \hspace{42pt} 0.5342 \\
\bottomrule 
\bottomrule                                                           
\end{tabular}}
\caption{nDCG@10 and MAP@100 for \rz with different first-stage retrieval models. For each, reranking is performed using the top 20 or 100 candidates.}
\label{tab:first_stages}
\end{table*}
\begin{table*}[t]
\centering \scalebox{0.75}{
\begin{tabular}{lllllll}
\toprule
\toprule
\multicolumn{1}{l}{\multirow{2}{*}{\hspace{120pt}}} & 
\multicolumn{2}{l}
{\multirow{2}{*}
{\begin{tabular}
[c]{@{}c@{}}\textbf{Source} \\
\textbf{Prev.} \hspace{10pt}  \textbf{Top-$k$}
\end{tabular}}} &
\multicolumn{2}{l}
{\multirow{2}{*}
{\begin{tabular}
[c]{@{}c@{}}\textbf{MSv2} \\
\textbf{DL21} \hspace{10pt}  \textbf{DL22}
\end{tabular}}} &
\multicolumn{2}{l}
{\multirow{2}{*}
{\begin{tabular}
[c]{@{}c@{}}\textbf{BEIR} \\
\textbf{NEWS} \hspace{5pt}  \textbf{COVID}
\end{tabular}}} \\
\multicolumn{1}{l}{}\ & \multicolumn{1}{l}{}\ & \multicolumn{2}{c}{} & \multicolumn{2}{c}{} \\
\midrule
(1) BM25                                       &
\multicolumn{2}{l}{None}{$|C|$}                 &
0.4458 & 0.2692                   &
0.3952 &  0.5947                 \\
(2) \rgthreefive                &
\multicolumn{2}{l}{BM25}{100}                           & 0.6050 & 0.4180 & 0.4936 & 0.7642  \\
(3) \rgfour                  &
\multicolumn{2}{l}{BM25}{100}                    & 0.7073 & 0.5078 & 0.5069 & 0.8471   \\
\midrule
(4) \rz & \multicolumn{2}{l}{BM25}{100} & 0.7029 & 0.5146 & 0.5184 & 0.8378 \\   
(5) \rzp & \multicolumn{2}{l}{BM25}{100} &0.7114 & 0.5164 & 0.5333 & 0.8377 \\ 
\bottomrule 
\bottomrule                                                           
\end{tabular}}
\caption{nDCG@10 on out-of-domain datasets:\ DL21 and DL22 from the \msvtwo (MSv2), and NEWS and COVID from the BEIR collection, extending \rz, \rzp, and \rgfour results from~\autoref{tab:1} for BM25. For each method, reranking is performed using the top 100 candidates from the previous stage.}
\label{tab:first_stages_dl21}
\end{table*}

\subsection{First-Stage Retrieval Model}
\label{sec:ablation_fstage}

This section delves into the influence of the quality and the quantity of candidates generated by the first-stage retriever on the final output.
To this end, our experiments employed four first-stage retrieval models, each with either the top 20 or top 100 results.
The methods include:\
(1) BM25~\cite{bm25},
(2) SPLADE++ EnsembleDistil (ED) \cite{spladepp},
(3) \repllama\cite{rankllama}, and 
(4) \adatwo~\cite{OpenAIEmbed, LuceneIsAll}. 
Results are presented in \autoref{tab:first_stages}.

Our findings clearly demonstrate that a better first-stage retrieval model leads to better overall reranking effectiveness. 
Specifically, stronger first-stage retrieval models such as SPLADE++ ED, rows~(2*), and \repllama, rows~(3*), markedly improve downstream effectiveness. 
We expect to see this as these methods yield better-quality candidate lists for reranking. 
However, a notable observation emerges when comparing the downstream effectiveness of reranking \repllama vs.\ reranking SPLADE++ ED. 
Despite \repllama exhibiting a higher MAP@100, reranking it does not translate to better results than reranking SPLADE++ ED, noticeable in DL19 and DL20 effectiveness, row~(2c) vs.\ (3c). 
While this gap can be partially ascribed to a 2\% reduction in the number of judged documents in the top 10 rankings of the \repllama reranked runs, there are likely additional factors as well.

We observe diminishing returns when \rz reranks more effective candidate lists.
For instance, while reranking the top 100 BM25 candidates with \rz results in a 45\%--55\% improvement in effectiveness across various metrics, the improvement is more modest with SPLADE++ ED, ranging from 7\%--20\% for the same metrics. 
This trend is consistent with observations in other multi-stage ranking systems~\cite{EMD, SqueezeMBERT, CT22}. 
A comparison between reranking the top 20 versus the top 100 candidates reveals that processing a larger pool generally leads to higher MAP@100 and, in most cases, also improves nDCG@10.
However, exceptions are noted in the DL19 dataset with dense retrievers, where reranking the top 20 yields better nDCG@10 scores than reranking the top 100. 
This is partially attributable to the 3\% difference in the percentage of the top 10 results judged.

\begin{table*}[ht]
\centering \scalebox{0.75}{
\begin{tabular}{lllllll}
\toprule
\toprule
\multicolumn{1}{l}{\multirow{2}{*}{\hspace{140pt}}} & 
\multicolumn{2}{l}
{\multirow{2}{*}
{\begin{tabular}
[c]{@{}c@{}}\textbf{Source} \\
\textbf{Prev.} \hspace{70pt}  \textbf{Top-$k$}
\end{tabular}}} &
\multicolumn{2}{l}
{\multirow{2}{*}
{\begin{tabular}
[c]{@{}c@{}}\textbf{DL19} \\
\textbf{nDCG@10} \hspace{20pt}  \textbf{MAP@100}
\end{tabular}}} &
\multicolumn{2}{l}
{\multirow{2}{*}
{\begin{tabular}
[c]{@{}c@{}}\textbf{DL20} \\
\textbf{nDCG@10} \hspace{20pt}  \textbf{MAP@100}
\end{tabular}}} \\
\multicolumn{1}{l}{} & \multicolumn{2}{l}{} & \multicolumn{2}{l}{} & \multicolumn{2}{l}{} \\
\midrule
(1a) \rgthreefive                               &
\multicolumn{2}{l}{BM25}{100}                   &
0.6855 & 0.3335             &
0.6202 & 0.3525            \\
(1b) \rgthreefive                               &
\multicolumn{2}{l}{Shuf.\ BM25}{100}            &
0.6158$\pm$0.012 & 0.2810$\pm$0.007             &
0.5516$\pm$0.011 & 0.2929$\pm$0.007             \\
(1c) \rgthreefive                               &
\multicolumn{2}{l}{SPLADE++ ED}{100}            &
0.7504 & 0.4731          &
0.7120 & 0.5011            \\
(1d) \rgthreefive                               &
\multicolumn{2}{l}{Shuf.\ SPLADE++ ED}{100}     &
0.6028$\pm$0.007 & 0.3403$\pm$0.005             &
0.5479$\pm$0.013 & 0.3500$\pm$0.009             \\
\midrule
(2a) \rz &
\multicolumn{2}{l}{BM25}{100} &
0.7420 & 0.3683 &
0.7086 & 0.4151 \\
(2b) \rz &
\multicolumn{2}{l}{Shuf.\ BM25}{100} & 0.7378$\pm$0.009 & 0.3421$\pm$0.008 & 0.7068$\pm$0.011 & 0.3910$\pm$0.007\\

(2c) \rz &
\multicolumn{2}{l}{SPLADE++ ED}{100} &
0.7816 & 0.4908 &
0.8159 & 0.5754 \\
(2d) \rz &
\multicolumn{2}{l}{Shuf.\ SPLADE++ ED}{100} & 0.7670$\pm$0.011 & 0.4287$\pm$0.005 & 0.7910$\pm$0.004 & 0.5112$\pm$0.004 \\
\bottomrule
\bottomrule
\end{tabular}}
\caption{nDCG@10 and MAP@100 on DL19 and D20 for \rgthreefive and \rz with different first-stage retrieval models.
For each, we perform reranking using the top 100 candidates from the previous stage on six shuffled orderings. 
We report average metrics with 99\% confidence intervals.}
\label{tab:shuffling}
\end{table*}

Additionally, we see that \rz reranking \repllama candidates, row~(3c), outperforms \rankllama's reranking of the same set; see \autoref{tab:1}, row~(3). 
This observation is particularly noteworthy considering the approximately 3\% fewer judged documents in the top 10 ranks in the case of \rz. 
These results imply that listwise rerankers might elevate a greater number of unjudged passages to higher ranks compared to pointwise methods.
This observation is possibly due to the historical dominance of pointwise methods in judgment pools and the relative paucity of listwise methods in 2019 and 2020.

To further explore this point, \autoref{tab:first_stages_dl21} presents additional results on out-of-domain datasets:\ DL21 and DL22 from the \msvtwo, and NEWS and COVID from BEIR.
Here, unlike~\autoref{tab:1}, rows (4), (7), and (8), we note that \rzp scores higher than \rgfour in three of the four tasks, rows (3), (4), and (5).
The lower effectiveness on COVID is unsurprising, especially given that the much larger \gptfour model is believed to have been exposed to large amounts of scientific text and thus holds more scientific knowledge.
\rzp improving over \rgfour is more important with BM25 as the first-stage retriever, given that we are reranking lower-quality candidates, making it imperative to promote poorly ranked relevant documents into top positions.

Summarizing, the impact of first-stage retrieval on overall effectiveness is substantial. 
The choice of retrieval model and the number of candidates considered for reranking are both crucial.
We also highlight the necessity of meticulously evaluating reranker models over different first-stage retrievers to capture nuances in a manner free of confounders.

\begin{table*}[t]
\centering \scalebox{0.75}{
\begin{tabular}{lccc}
\toprule
\toprule
\multicolumn{1}{l}{\multirow{2}{*}{\hspace{100pt}}} &  
\multicolumn{3}{c}{
\multirow{2}{*}{
\begin{tabular}{@{}c@{}}
\textbf{NovelEval} \\
\textbf{nDCG@1} \hspace{18pt} \textbf{nDCG@5} \hspace{20pt} \textbf{nDCG@10}
\end{tabular}}} \\
\\
\midrule
(1) BM25 & \hspace{8pt} 0.3333 & \hspace{25pt} 0.4596 & \hspace{19pt} 0.5577 \\
(2) monoT5 (3B) & \hspace{8pt} 0.8333 & \hspace{25pt} 0.7838 & \hspace{19pt} 0.8462 \\
\midrule
(3) \rgthreefive & \hspace{8pt} 0.7619 & \hspace{25pt} 0.7415 & \hspace{19pt} 0.7571 \\
(4) \rgfour &  \hspace{8pt} 0.8571 & \hspace{25pt} 0.8749 & \hspace{19pt} 0.9045 \\
\midrule
(5) \rz & \hspace{8pt} 0.9286 & \hspace{25pt} 0.8615 & \hspace{19pt} 0.8934\\
\bottomrule 
\bottomrule
\end{tabular}}
\caption{nDCG@\{1,5,10\} on the NovelEval-2306 set~\cite{RankGPT} for different reranking methods with Google Search as the first stage. Each reranker uses the top 20 retrieved results of the previous stage as input.}
\label{tab:novel-eval}
\end{table*}

\subsection{Data Augmentation}

This section seeks to better understand the robustness benefits of data augmentation (DA) in enhancing model effectiveness, especially in the presence of distractors (shuffled orderings), as suggested by~\citet{RankVicuna}, within the context of our newly developed \rz models.

The training methodology of \rz (see Section~\ref{sec:rankzephyr-stage2}) includes sequences formed from altering the input order of documents and accordingly modifying the teacher generations. 
This process presents a more complex reordering challenge to the model, contributing to its robustness and effectiveness.
Our investigation examines whether the effectiveness improvements observed in previous open-source LLMs persist in \rz and how it compares with \rgthreefive.
We present our experimental findings in \autoref{tab:shuffling}.

An examination of rows~(2a) and (2b), as well as (2c) and (2d), reveals that \rz retains its effectiveness with shuffled inputs. 
A minor decrease in effectiveness is expected, possibly attributable to the model's inability to simultaneously promote multiple relevant documents in cases where they are initially ranked lower, such as in ranks 81--100.
Such occurrences are relatively rare but can actually appear anywhere in the ranked lists.

Conversely, \rgthreefive demonstrates a marked decrease in effectiveness when fed shuffled inputs, as indicated by the contrast between rows~(1b) and (1a), and (1d) and (1c).
Notably, the effectiveness of \rgthreefive in reranking shuffled SPLADE++ ED is even lower than that of reranking shuffled BM25.
This pronounced dependence on input order underscores a lack of robustness.

In summary, our analysis highlights the robustness of \rz to input perturbations, reinforcing the value of data augmentation during model training.
Given that even good retrievers might produce sub-optimal rankings that behave like shuffled input, it is vital to build methods like \rz or corrective strategies like \rzp to ensure both effectiveness and robustness.

\subsection{Uncontaminated Data: NovelEval}

Given that many widely used benchmark datasets today were created years ago, there is a possibility that LLMs have already memorized knowledge specific to those queries due to contamination of the training (or worse, test) data.
To address this possibility, \citet{RankGPT} proposed NovelEval, a continuously updated retrieval test set, ensuring that queries and candidate passages have not been seen during training.

\autoref{tab:novel-eval} presents our results on NovelEval-2306, a test set with 21 questions from four domains, curated in June 2023.
\citet{RankGPT} also provided the top 20 candidates from Google Search that have been manually judged on a scale from 0 to 2.
Since we are reranking only the top 20 results, which is the size of a single window, we do not employ progressive reranking.
Additionally, we report metrics in the order the authors presented them, focusing on nDCG@1 because of their emphasis on high precision.
Row~(1) shows the results of BM25; row~(2), monoT5 (3B)~\cite{monoT5}, an effective pointwise reranker based on T5; rows (3--4),  \rgthreefive and \rgfour, and row~(5), \rz.

We find that in terms of the primary metric, \rz outperforms \rgthreefive and \rgfour, 
row~(5) vs.\ (3) and (4).
This suggests that in a RAG setting, where the model uses only the top candidate, \rz provides a great alternative to proprietary models like \rgfour.
Looking at the other metrics, we see that \rz still, for the most part, bridges the gap with \rgfour.

\section{Conclusions}

In this paper, we present \rz, a novel open-source LLM optimized for zero-shot listwise 
 reranking.
Our extensive experiments demonstrate that \rz matches and occasionally surpasses the effectiveness of much larger proprietary models such as \rgfour. 
The significance of our work lies not just in reranking effectiveness, but also in reproducibility and accessibility, setting a new standard for the community.

Our work delves into multiple facets of reranking with prompt-decoder models.
We show that progressive reranking generally yields higher-quality output in being able to promote more relevant documents.
We also show how the choice of the first-stage retrieval model impacts downstream reranking effectiveness in \rz and other listwise methods. 
We explored settings in the teacher model, candidate source, and sampling strategies that lead to the most efficient and effective instruction fine-tuning.
Our findings also highlight the crucial role of data augmentation in enhancing robustness. 
Finally, the effectiveness of \rz on the NovelEval test set, comprising ``uncontaminated'' queries and passages, addresses potential concerns about memorization and potential for real-world applications.

We believe that \rz represents a noteworthy milestone as the community continues to develop and refine large language models for information retrieval.
Our work paves the way for future exploration in more effective and robust reranking models, both as an independent component and as part of a larger RAG pipeline.

\section*{Acknowledgments}

This research was supported in part by the Natural Sciences and Engineering Research Council (NSERC) of Canada.
Computational resources were provided in part by Microsoft in the form of Azure credits as part of the ``Accelerating Foundation Models Research'' program.

\balance

\bibliography{custom}
\bibliographystyle{acl_natbib}

\end{document}